\documentclass[%
 reprint,
superscriptaddress,
showpacs,
 amsmath,amssymb,
 aps,
]{revtex4-1}

\usepackage{graphicx}
\usepackage{color}
\usepackage{verbatim}

\begin{document}

\preprint{APS/123-QED}

\title{Engineering continuous and discrete variable quantum vortex states by\\ nonlocal photon subtraction in a reconfigurable photonic chip}

\author{David Barral}\email{Corresponding author. Email: david.barral@usc.es}
\author{Jes\'us Li$\tilde{\rm{n}}$ares}
\author{Daniel Balado} 
\affiliation{Optics Area, Department of Applied Physics, Faculty of Physics and Faculty of Optics and Optometry, University of Santiago de Compostela, Campus Vida s/n (Campus Universitario Sur), E-15782 Santiago de Compostela, Galicia, Spain.}

\begin{abstract} We study the production of entangled two- and N-mode quantum states of light in optical waveguides. To this end, we propose a quantum photonic circuit that produces a reconfigurable superposition of photon subtraction on two single-mode squeezed states. Under postselection, continuous variable or discrete variable entangled states with possibilities in quantum information processing are obtained. Likewise, nesting leads to higher-dimension entanglement with a similar design, enabling the generation of non-Gaussian continuous variable cluster states. Additionally, we show the operation of the device through the generation of quantum vortex states of light and propose an integrated device that measures their order and handedness. Finally, we study the non-Gaussianity, nonclassicality, and entanglement of the quantum states generated with this scheme by means of the optical field- strength distribution, Wigner function, and logarithmic negativity.
\end{abstract}



\maketitle 

\section{Introduction}

Continuous-variable encoded quantum information processing (CV-QIP) has emerged in recent years as a solid alternative to its discrete-variable counterpart (DV-QIP). Continuous-variable entanglement is the key element for a great number of quantum information protocols \cite{Braunstein2005}. These algorithms have been mostly carried out with Gaussian entangled states, but demonstrations of the inability of these states to perform entanglement distillation protocols under certain circumstances have been shown \cite{Eisert2002, Fiurasek2002, Giedke2002}. This has led to the study of non-Gaussian states overcoming these drawbacks by means of non-Gaussian operations \cite{Takahashi2010}, a prominent example of photon-level quantum light engineering (QLE), which has attracted great attention over the last years \cite{Kim2008}. QLE has interesting applications both in QIP as well as in fundamental quantum optics. Manipulating light at the photon level by means of subtraction and addition provides a simple way to construct these non-Gaussian states, like the creation of quantum states from classical states by adding a single photon \cite{Agarwal1991, Zavatta2004} or the generation of Schr\"{o}dinger-cat-like states through conditional absorption of photons on a squeezed vacuum \cite{Dakna1997, Neergaard2006, Ourjoumtsev2007}, as well as testing the foundations of quantum mechanics, like the bosonic commutation relation by means of the coherent superposition of photon addition followed by photon subtraction (and vice versa) \cite{Kim2008b, Zavatta2009} or carrying out loophole-free Bell tests \cite{Nha2004, GarciaPatron2004}. What is more, the application of this kind of states has been recently shown in CV quantum teleportation, showing the transfer of the non-Gaussianity of the state \cite{Lee2011}.

Likewise, in recent times the use of the orbital angular momentum (OAM) spatial degree of freedom of optical vortex fields as a quantum information resource has yielded ground-breaking demonstrations in QIP \cite{Molina2007}. In this regard, there is a class of two-mode non-Gaussian states with interesting related features: the quantum vortex states \cite{Agarwal1997}. These are entangled squeezed eigenstates of the $z$ component of the abstract angular momentum operator $\hat{L}_{z}$, in analogy with the quantum eigenstates of the spatial OAM operator, which show interesting non-classical properties. Other families of states have been named quantum vortices because of its similarity with spatial-mode optical vortices: SU(2)-transformed Fock states \cite{Agarwal2006}, generalized quantum vortices \cite{Bandyopadhyay2011, Bandyopadhyay2011b, Agarwal2011, Banerji2014}, Bessel-Gauss quantum vortex states \cite{Zhu2012, Zhu2012b} or Hermite polynomial quantum vortices \cite{Li2015}. These are strongly entangled states \cite{Agarwal2006, Agarwal2011}, being therefore interesting for CV-QIP. In this direction, Li {\it et al.} have lately shown their performance in a teleportation protocol \cite{Li2015}. Moreover, quantum vortices have recently received attention in the context of quantum polarization and Bohmian dynamics \cite{Luis2013, Luis2015}.

\begin{figure*}[t]
\centering
\includegraphics[width=\textwidth]{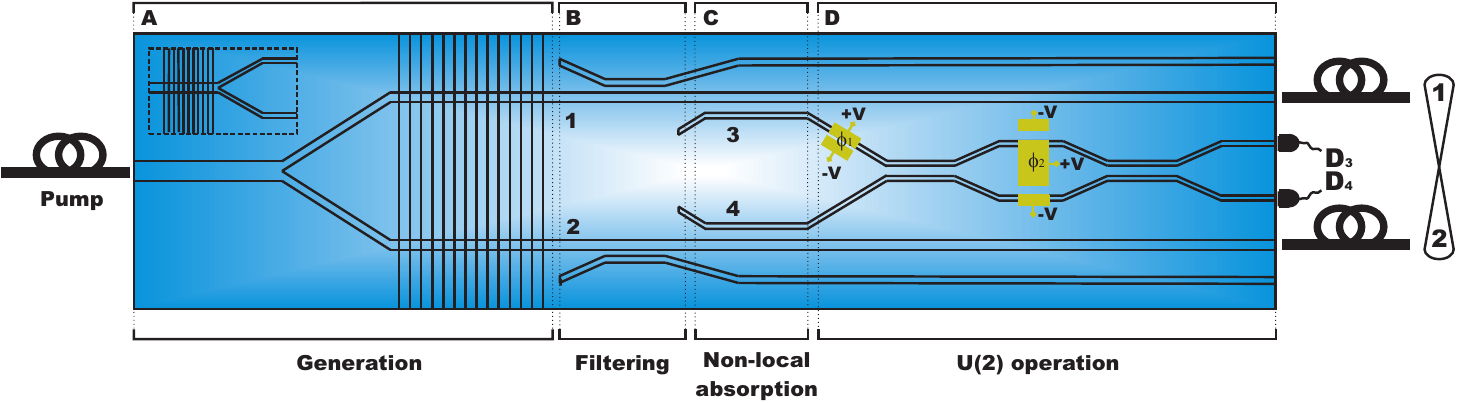}
\vspace {0cm}\,
\hspace{0cm}\caption{\label{F1}\small{(Color online) Sketch of the photonic circuit proposed for engineering CV entangled quantum states. The first stage is devoted to the generation of the quantum state (A) and pump suppression (B). The production of two single-mode squeezed states is carried out by means of a periodically poled zone. In the inset an alternative design of the generation stage is shown (dashed-line square). In the second stage a pair of directional couplers weakly coupled carry out non-local absorption of photons (C). Finally, in the third stage an U(2) operation on these photons is performed by means of a reconfigurable Mach-Zehnder interferometer (D). Measurement of these photons leads to the entanglement of the input factorable squeezed states.}}
\end {figure*}

On the other hand, integrated photonics has led to multiple new approaches on QIP with high success \cite{Tanzilli2012}. These are based on the processing of quantum states of light in photonic circuits, either DV or CV-encoded \cite{OBrien2009}, and their measurement by single-photon detectors or homodyne detection schemes, respectively \cite{Silberhorn2007}. The main features for which integrated optics circuits are crucial to QIP and quantum optics are the sub-wavelength stability and the great miniaturization they show with respect to bulk optics analogs, providing high-visibility quantum interference and scalability, indispensable as the level of complexity of the circuit increases; as well as the optical properties of the materials which make up the waveguides, enabling the generation of quantum states on chip by means of their enhanced nonlinear features \cite{Rogers2015, Dutt2015}, the manipulation by means of their thermo-optic and electro-optic properties \cite{Jin2014, Masada2015, Carolan2015}; and even the integration of detectors in the circuit \cite{Sahin2015, Najafi2015}. This leads to high integration density, fidelity and fast control. Therefore, this technology has the potential to turn the real power of quantum mechanics into reality with practical, low cost, standardized, interconnectable and reconfigurable components \cite{Tanzilli2012}.

Bearing all the above in mind, the motivation of this article is to design a quantum circuit on-chip able to generate and manipulate CV entangled states of light by means of directional couplers, phase shifters and conditional measurement; and to show its performance with quantum vortex states. As we will show, entanglement is created by non-local photon absorption of two separable states by means of directional couplers with high transmittivity. This delocalized photon is locally manipulated by means of a reconfigurable interferometer which produces the desired state after photon counting. The quantum states thus produced hybridize appealing features of both CV (squeezed light) and DV (single photons) \cite{Andersen2015}. Likewise, the integrated nature of this circuit provides access to entanglement in higher-dimension Hilbert spaces via nesting, with application in measurement-based quantum computing \cite{Andersen2015}. Furthermore, we demonstrate the versatility of this design as it can also be used to generate DV entangled states by only changing the power of the input pump in such a way that just one monolithic chip can be used in different QIP tasks. Finally, we introduce a possible application of this device by means of the generation of CV and DV quantum vortex states. To the best of our knowledge, this is the first proposal for the generation and manipulation of this family of states in photonic circuits, presenting also an integrated detection scheme able to detect the order and handedness of the quantum vortices. We study quantum features of these states like non-Gaussianity, non-classicality and entanglement by means of their associated optical field-strength distribution, Wigner function and logarithmic negativity.

The article is organized as follows: Section II presents a scheme of generation of general CV and DV entangled quantum states in a reconfigurable chip, focusing afterwards on quantum vortices and studying their field-strength probability and phase distributions, as well a discussion about the merits of this proposal. Section III is devoted to study the non-classicality and entanglement through the Wigner distribution and logarithmic negativity of the generated states. 
Finally, the main results of this article are summarized in Section IV. 

\section{CV \& DV entanglement in a reconfigurable photonic chip}

\subsection{Device operation}

Our QLE-device is pictured in Figure \ref{F1}. It is made up of three stages. The first one (Figure \ref{F1}A) is devoted to the generation of two single-mode squeezed states. There are different approaches in integrated optics with the ability to produce these states \cite{Tanzilli2012}. One of the more widespread is the use of periodically poled waveguides. Tailoring the nonlinearity of the material, usually lithium niobate (PPLN) or potassium titanyl phosphate (PPKTP), enables quasi-phase matching of the propagation constants (QPM) improving the efficiency of conversion between the pump and the the desired quantum state \cite{Tanzilli2000}. First order QPM is given by
\begin{equation} \label{QPM}
\Delta\beta\equiv\beta(\omega_{p})-\beta(\omega_{s})-\beta(\omega_{i})=\frac{2\pi}{\Lambda},
\end{equation}
where $\Delta\beta$ represents the propagation constants mismatch between a pump photon (p) and its daughters photons, labelled signal (s) and idler (i), caused by dispersion; and $\Lambda$ is an appropriate period of the inverted domains. This has been shown in the generation of both entangled twin photons \cite{Suhara2009} and squeezed states \cite{Yoshino2007} on chip. Likewise, generation of entanglement in different degrees of freedom, as mode number, path, frequency or polarization, as well as hyperentanglement has been proposed in recent years through domain engineering on periodically poled waveguides \cite{Saleh2009, Lugani2010}. 

Two different possibilities are sketched in Figure \ref{F1}A. The main one shows a two-mode waveguide pumped with a single-mode coherent field followed by a symmetric $Y$ splitter and two separated waveguides with a common periodic poled grating in the substrate, where the period is tailored such that type-I {spontaneous parametric down-conversion} (SPDC) is generated \cite{Jin2014}. The generation of quantum light in waveguides can be mathematically represented by the Momentum operator $\hat{M}$ \cite{Linares2008, Barral2015a}. In this case, the corresponding Momentum related to the interacting optical fields (interaction picture) for each waveguide $j=1, 2,$ is given by
\begin{equation} \label{M1}
\hat{M}_{j}=\frac{1}{2} \iiint \epsilon_{0} \chi_{eff}^{(2)} (\sum_{k} \hat{E}_{j ,k})^{2}  \,dx dy dt,
\end{equation}
with $k=i, s, p$ and $\epsilon_{0}$ is the vacuum permittivity, $\chi_{eff}^{(2)}$ the second order effective nonlinearity in the poling area and the integral is performed over the transverse area of the waveguide and the period of the waves. Likewise, the quantum optical fields are given by
\begin{equation}
\hat{E}_{j, k}\propto \hat{a}_{j, k} \,e^{i \beta_{j}(\omega_{k}) z} \, e^{-i\omega_{k} t} \,{\mathbf e_{j, k}(x,y)} + h.c.,
\end{equation}
with $\hat{a}_{j, k}$ and ${\mathbf e_{j, k}(x,y)}$ the absorption operators and the normalized transverse vector amplitudes related to each mode $j, k$, respectively. Then, pumping with a strong coherent field and taking into account the conservation of energy and the QPM, in such a way that the signal and idler waves are degenerated in frequency $(\omega_{p}/2=\omega_{s}=\omega_{i}$), we obtain the following Momentum after applying the above fields into Equation $(\ref{M1}$) and carrying out the integrals \cite{Linares2008}
\begin{equation} \label{M2}
\hat{M}_{j}^{I}(\Gamma)=-\frac{ i \hbar}{2} (\Gamma \hat{a}_{j}^{\dag 2} - \Gamma^{*} \hat{a}_{j}^{2}),
\end{equation}
with $\Gamma$ a nonlinear coupling constant which depends on the pump intensity, the effective nonlinearity of the material $\chi_{eff}^{(2)}$, the mode mismatch and the QPM. 

On the other hand, the inset in Figure \ref{F1}A sketches a second possibility: a two-mode waveguide with a periodic poled grating satisfying simultaneously QPM for the two modes in a type-0 SPDC followed by a symmetric $Y$ junction. In this case, a coherent pump excited in the odd spatial mode produces a two-mode squeezed vacuum excited in both the even and odd spatial modes of the waveguide which are set to be degenerated in frequency. The selection of the pump mode  as odd is related to the overlap integral of the transverse profiles of the interacting modes, which would be zero if the pump mode is even \cite{Saleh2009}. The excitation of the odd mode of this two-mode waveguide could be carried out by means of a binary phase plate transforming the spatial structure of the pump from a Gaussian to a first-order Hermite-Gaussian mode, before entering the waveguide \cite{Bai2012}. The Momentum operator related to this two-mode periodically poled waveguide is then given by
\begin{equation} \label{M3}
\hat{M}^{II}=-i \hbar(\Gamma' \hat{a}_{o}^{\dag} \hat{a}_{e}^{\dag} - \Gamma'^{*} \hat{a}_{o}  \hat{a}_{e} ),
\end{equation}
where $o$ and $e$ stand here for the first (even) and second (odd) excited guided modes. The light finds then the symmetric $Y$ junction afterwards. This device performs the following transformation
\begin{equation} \label{Yjunction}
 \begin{pmatrix}
 \hat{a}_{o} \\
 \hat{a}_{e}
 \end{pmatrix}
 \approx \frac{1}{\sqrt{2}}
  \begin{pmatrix}
     1 &  -1 \\
     1 &  1
  \end{pmatrix} 
    \begin{pmatrix}
 \hat{a}_{1} \\
 \hat{a}_{2}
  \end{pmatrix} , 
\end{equation}
where we have used the approximation since part of the light is lost in the junction in the form of radiation modes. Applying Equation (\ref{Yjunction}) into Equation (\ref{M3}), the next Momentum operator is obtained
\begin{equation}
\hat{M}^{II} \rightarrow \sum_{j=1,2}\hat{M}_{j}^{I}(\Gamma').
\end{equation}
It should be outlined that this generation scheme is less efficient than the first one, since the spatial overlap between the the pump and the SPDC modes is not as good as in the first case, due to their different spatial parity. Likewise, in this case part of the entangled quantum states generated in the PPLN zone would be lost at the $Y$ junction, lowering the flux of quantum light. Both schemes lead then to two degenerated single-mode squeezed vacuum states spatially separated, each one given by the following propagation operator
\begin{equation}\label{SMSO}
\hat{S}_{j}(\zeta)=e^{\frac{i}{\hbar} \hat{M}_{j} L_{p} }=e^{\frac{1}{2}(\zeta \hat{a}_{j}^{\dag 2} - \zeta^{*} \hat{a}_{j}^{2})},
\end{equation}
where $L_{p}$ is the length of the poling area and $\zeta \equiv r e^{i\theta_{s}}=\Gamma L_{p}$ (or $\Gamma' L_{p}$) is the complex squeezing parameter generated in the poling area. We outline here that we have chosen above type-I and type-0 SPDC processes in order to have the same polarization in both the signal and idler guided modes.

Moreover, the pump can be filtered by means of properly designed wavelength dependent directional couplers (DC) \cite{Kanter2002, Jin2014}, as shown in Figure \ref{F1}B. This operation is given in terms of operators by  
\begin{equation}\label{DC}
\hat{B}_{j\,l} (\theta_{j}(\omega))=e^{-i (\theta_{j}(\omega)/2) \hat{\sigma}_{x}},
\end{equation}
where we have defined $\theta_{j} (\omega)\equiv 2\kappa(\omega)L_{F}$ and $\kappa(\omega)$ is the frequency-dependent coupling strength of the directional coupler of length $L_{F}$ and $\hat{\sigma}_{x}=\hat{a}_{j} \hat{a}_{l}^{\dag}+\hat{a}_{j}^{\dag} \hat{a}_{l}$ is the corresponding Pauli operator. $\hat{B}_{j\,l}$ stands for a directional coupler with reflectivity $r_{j}=\sin(\theta_{j}(\omega)/2)$ and transmittivity $t_{j}=\cos(\theta_{j}(\omega)/2)$. In our case these devices are designed such that they fully reflect the pump ($\theta(\omega_{p})=\pi$) towards the ancillary waveguides $l$ and transmit the squeezed vacuum ($\theta(\omega_{s})=2\pi$) through the signal waveguides $j=1, 2$. Therefore, we obtain after this stage the following factorable two single-mode squeezed vacuum state
\begin{equation} \label{TwoSingleSq}
\vert \Psi_{B} \rangle = \hat{S}_{1} (\zeta) \hat{S}_{2} (\zeta) \vert 0_{1} \, 0_{2} \rangle \equiv  \vert 0_{\zeta\,1} \, 0_{\zeta\, 2} \rangle,
\end{equation}
where $\vert 0_{\zeta} \rangle$ accounts for the single-mode squeezed vacuum.

The second stage (Figure \ref{F1}C) is comprised of two directional couplers with high transmittivity, where a small fraction of each mode is reflected and sent to the third stage. The mixing of modes $1$ and $3$ on one hand and $2$ and $4$ on the other, produces the following state
\begin{equation}
\vert \Psi_{C} \rangle = \hat{B}_{1 3} (\theta_{1}) \hat{B}_{2 4} (\theta_{2}) \vert 0_{\zeta \,1} \, 0_{\zeta \, 2} \, 0_{3} \, 0_{4} \rangle,
\end{equation}
where $\theta_{j}\equiv\theta_{j}(\omega_{s})$.
As input modes $3$ and $4$ are in the vacuum, using the disentangling theorem for the SU(2) group \cite{Kim2008}, we can write
\begin{equation}
\begin{split}
\vert \Psi_{C} \rangle = e^{-i \tan(\theta_{1} /2) \hat{a}_{1} \hat{a}_{3}^{\dag}} &\,t_{1}^{\hat{a}_{1}^{\dag} \hat{a}_{1}} \,e^{-i \tan(\theta_{2} /2) \hat{a}_{2} \hat{a}_{4}^{\dag}} \,t_{2}^{\hat{a}_{2}^{\dag} \hat{a}_{2}} \\
&\times \vert 0_{\zeta \,1} \, 0_{\zeta \, 2} \, 0_{3} \, 0_{4} \rangle,
\end{split}
\end{equation}
which under the conditions of high transmittivity $t_{1,2}\approx1$ and moderate squeezing can be approximated by \cite{Kim2008b}
\begin{equation}
\vert \Psi_{C} \rangle = (1-i\frac{r_{1}}{t_{1}} \hat{a}_{1} \hat{a}_{3}^{\dag})(1-i\frac{r_{2}}{t_{2}} \hat{a}_{2} \hat{a}_{4}^{\dag}) \vert 0_{\zeta \,1} \, 0_{\zeta \, 2} \, 0_{3} \, 0_{4} \rangle,
\end{equation}
where single photons propagating through the lossy channels $3$ and $4$ appear with a certain probability. 

The last stage is an integrated Mach-Zehnder interferometer (MZI) composed of two $3$ $dB$ directional couplers with an active phase shifter in between, preceded by another phase shifter, with the modes $3$ and $4$ as input states (Figure \ref{F1}D). This device performs any U(2) operation on the input states adjusting the values of $\phi_{1}$ and $\phi_{2}$, being these the phases generated by the two phase shifters, respectively. The fidelity of this kind of devices has been thoroughly demonstrated along the last decade with both electro-optic and thermo-optic material supports \cite{Martin2012, Bonneau2012, Shadbolt2012, Carolan2015}. If the material support selected is lithium niobate, considering the waveguides surrounded by electrodes, the phase shift is given by $\phi=n^{3} r V\pi L_{E}/ \lambda d$, with $n$ the ordinary or extraordinary refractive index on the material for a given wavelength $\lambda$, depending on the input polarization, $r$ the relevant component of the second order nonlinear tensor, $V$ the applied voltage, $L_{E}$ the length of the electrodes and $d$ the distance between them. In the case of the MZI, two reversed electrodes are used in order to lower the required voltage by a factor of two compared with an only phase shifter. A switching efficiency of $\approx98\%$ and switching times of $4$ ns have been reported with a MZI configuration \cite{Bonneau2012}. Likewise, this operation could be carried out with an alternating $\Delta\beta$ directional coupler, which could improve the integration density and fidelity of the operation, as proposed in \cite{Barral2015}. Then, applying this transformation on $\vert \Psi_{C} \rangle$ we have the following state \cite{Campos1989}
\begin{equation} \label{GenState}
\begin{split}
\vert &\Psi_{D} \rangle= \hat{U}_{3 4} (\phi_{1}, \phi_{2}) \vert \Psi_{C} \rangle \\
&=\{1-i\frac{r_{1}}{t_{1}} e^{i \phi_{1}/2}  \hat{a}_{1} [ \cos(\phi_{2}/2) \hat{a}_{3}^{\dag} + \sin(\phi_{2}/2) \hat{a}_{4}^{\dag} ] \\
&-i\frac{r_{2}}{t_{2}} e^{-i \phi_{1}/2}  \hat{a}_{2} [ -\sin (\phi_{2}/2)  \hat{a}_{3}^{\dag} + \cos(\phi_{2}/2)  \hat{a}_{4}^{\dag}] +\mathcal{O}(r^{2}) \} \\
&\times \vert 0_{\zeta \,1} \, 0_{\zeta \, 2} \, 0_{3} \, 0_{4} \rangle,
\end{split}
\end{equation}
where $\hat{U}_{3 4}(\phi_{1}, \phi_{2})=e^{i\phi_{2}\hat{\sigma}_{y}/2}\,e^{i\phi_{1} \hat{\sigma}_{z}/2}$ and $\hat{\sigma}_{y}$ and $\hat{\sigma}_{z}$ are the well known Pauli operators. The terms quadratic in the reflectivity are neglected since large transmittivity is considered. 

\begin{figure}[t]
\centering
\includegraphics[width=0.47\textwidth]{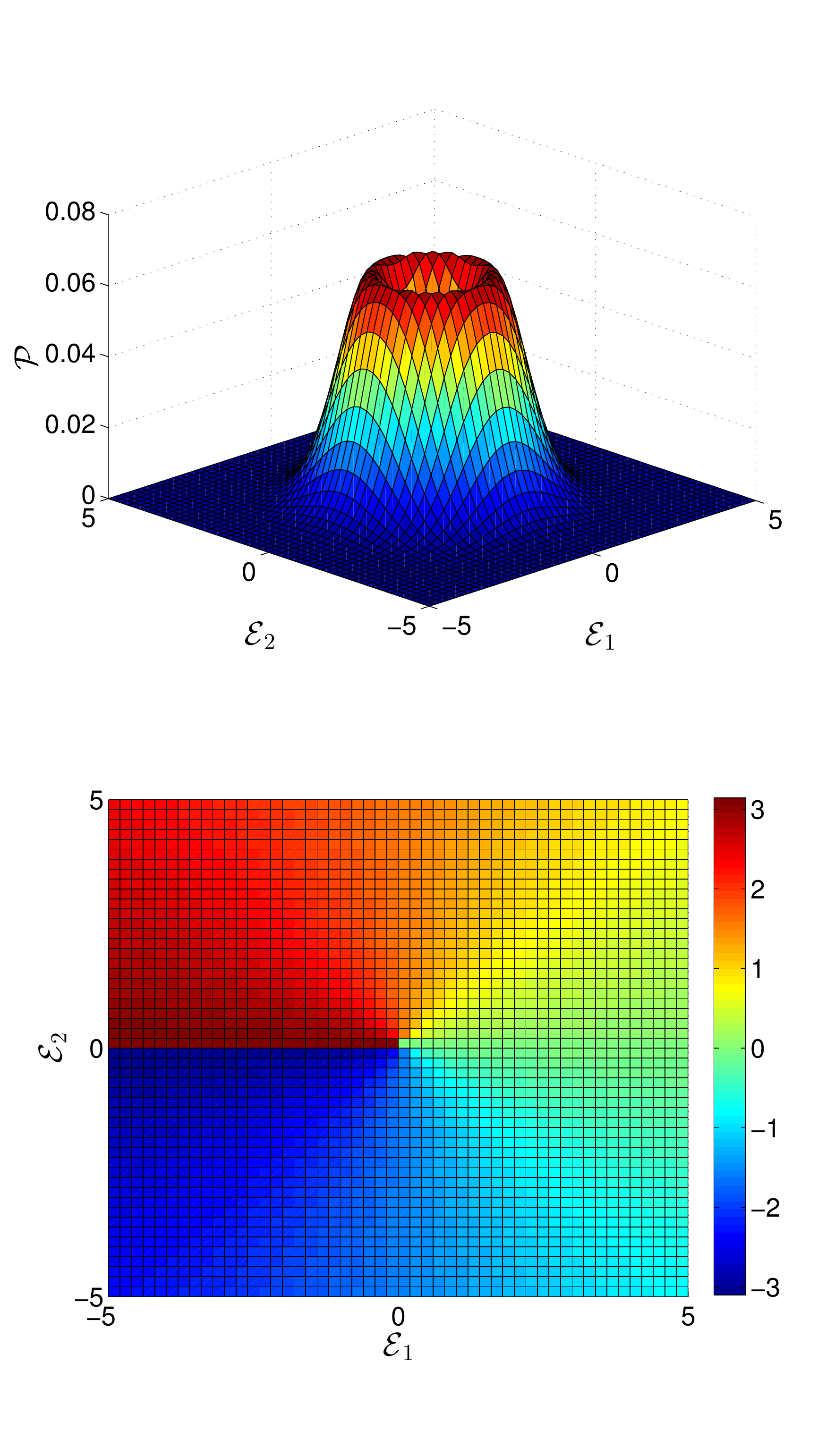}
\vspace {0cm}\,
\hspace{0cm}\caption{\label{F2}\small{(Color online) Plots of probability (upper figure) and phase (lower figure) densities of a CV circular quantum vortex with squeezing factor $r=0.3$ and $\eta'=1$.}}
\end {figure}


\begin{figure}[h]
\centering
\includegraphics[width=0.47\textwidth]{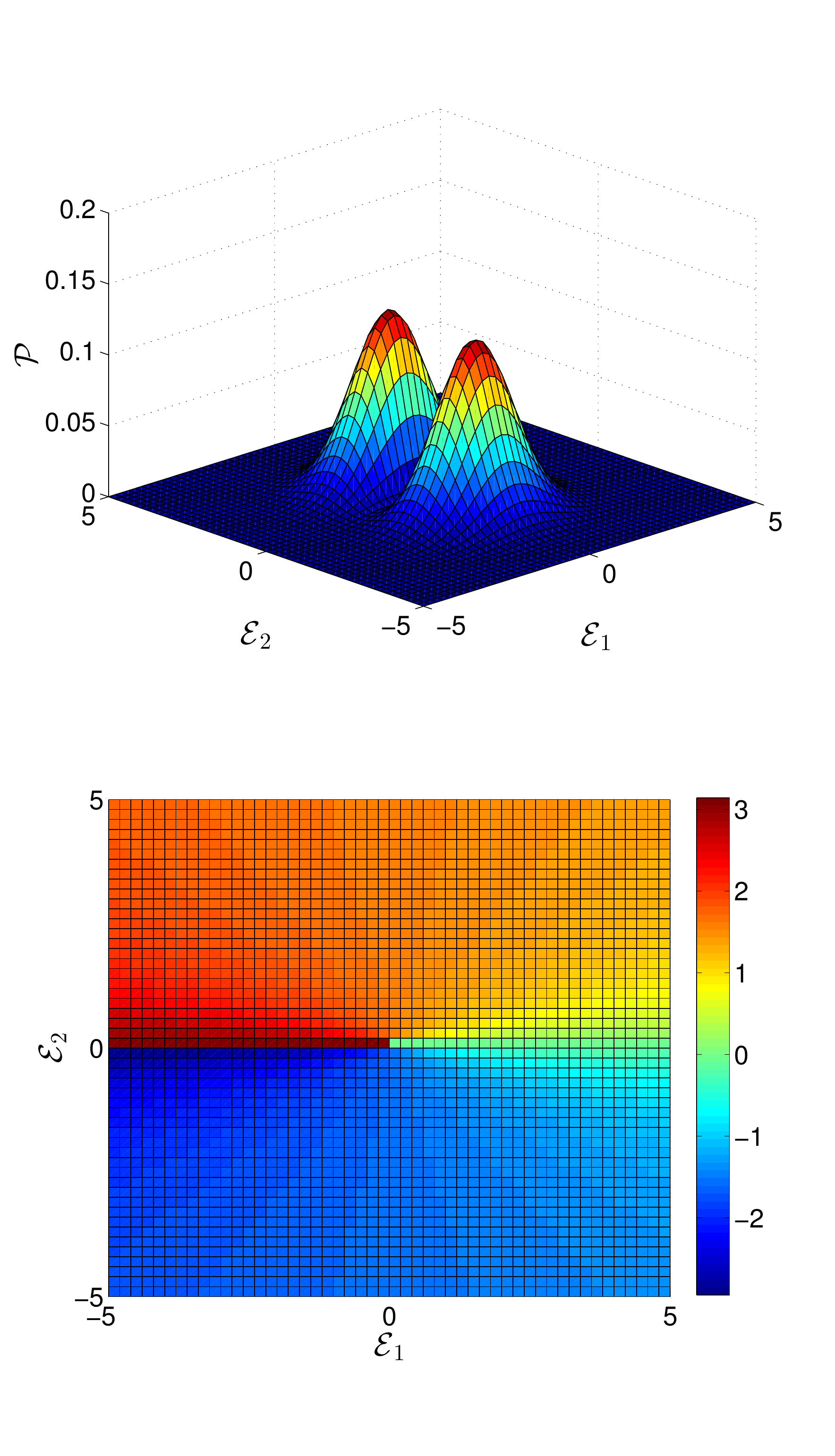}
\vspace {0cm}\,
\hspace{0cm}\caption{\label{F3}\small{(Color online) Plots of probability (upper figure) and phase (lower figure) densities of a CV elliptical quantum vortex with squeezing factor $r=0.3$ and $\eta'=5$.}}
\end {figure}

\subsection{CV quantum vortex states}

Next, we present a post-selection method to generate non-Gaussian CV entangled states.
It is easy to see that the detection of a photon in the modes $3$ or $4$ will herald the subtraction of a delocalized photon, generating the following entangled states
\begin{subequations} \label{CVeqs}
\begin{equation}
\begin{split}
&\langle 1_{3} 0 _{4}\vert \Psi_{D} \rangle \propto \\
&[\frac{r_{1}}{t_{1}} \cos(\phi_{2}/2) \hat{a}_{1} - \frac{r_{2}}{t_{2}}  \sin (\phi_{2}/2) e^{-i\phi_{1}} \hat{a}_{2} ] \vert 0_{\zeta \,1} \, 0_{\zeta \, 2} \rangle, 
\end{split}
\end{equation}
\begin{equation}
\begin{split}
&\langle 0_{3} 1_{4}\vert \Psi_{D} \rangle \propto  \\
&[\frac{r_{1}}{t_{1}} \sin(\phi_{2}/2) e^{i\phi_{1}} \hat{a}_{1} + \frac{r_{2}}{t_{2}} \cos (\phi_{2}/2) \hat{a}_{2} ] \vert 0_{\zeta \,1} \, 0_{\zeta \, 2} \rangle. 
\end{split}
\end{equation}
\end{subequations}
These measurements can be carried out with single-photon avalanche photodiodes (SPADs) or superconducting single photon detectors (SSPDs), inter alia \cite{Eisaman2011}. Moreover, taking into account the single-mode squeezing Bogoliubov transformation \cite{Agarwal2012}
\begin{equation} \label{Bogoliubov}
\hat{S}^{\dag}(\zeta) \hat{a} \hat{S}(\zeta)= \hat{a} \cosh(r) + \hat{a}^{\dag} \sinh(r)\,e^{i\theta_{s}},
\end{equation}
we can write Equations (\ref{CVeqs}) as follows:
\begin{subequations} \label{SVS}
\begin{equation}\label{SVSa}
\langle 1_{3} 0_{4} \vert \Psi_{D} \rangle \propto  \vert 1_{\zeta 1} 0_{\zeta 2} \rangle -\eta \,\tan(\phi_{2}/2) e^{-i\phi_{1}} \vert 0_{\zeta 1} 1_{\zeta 2} \rangle ,            
\end{equation}
\begin{equation}\label{SVSb}
\langle 0_{3} 1_{4}\vert \Psi_{D} \rangle \propto \vert 1_{\zeta 1} 0_{\zeta 2} \rangle +\eta \,\cot(\phi_{2}/2) e^{-i\phi_{1}} \vert 0_{\zeta 1} 1_{\zeta 2} \rangle,
\end{equation}
\end{subequations}
with $\eta=\frac{r_{2} t_{1}}{r_{1} t_{2}}$ and where $\vert 1_{\zeta} \rangle=\hat{S}(\zeta) \hat{a}^{\dag}\vert 0 \rangle$ accounts for a squeezed single photon. Both states are two-mode strongly entangled states, as we will see in the next section. Fixing the phase $\phi_{1}$ to any value $(2n-1) \pi/2$ we obtain an elliptical vortex state, with a degree of ellipticity dependent on the values of $\phi_{2}$ and $\eta$. Focusing on Equation (\ref{SVSa}), we can write the elliptical quantum vortex as
\begin{equation}\label{SVSE}
\vert \Psi \rangle = \frac{1}{\sqrt{1+\eta'^{2}}} ( \vert 1_{\zeta 1} 0_{\zeta 2} \rangle + (-1)^{n} i\,\eta' \vert 0_{\zeta 1} 1_{\zeta 2} \rangle) ,            
\end{equation}
where we have defined $\eta'=\eta \,\tan(\phi_{2}/2)$ as an ellipticity factor. It is important to outline that this approach enables the manipulation of the quantum state as well as correct fabrication errors or asymmetries between the couplers $B_{1\,3}$ and $B_{2\,4}$, represented by values of $\eta\neq1$. Likewise, adjusting the phase of the second phase shifter such that $\phi_{2}=2 \arctan(\eta^{-1})$ (or $\phi_{2}=2 \arctan(\eta) $ in Eq. (\ref{SVSb})) we obtain the vortex state of the quantized optical field or circular quantum vortex \cite{Agarwal1997}
\begin{equation}\label{SVSC}
\vert \Psi \rangle =\frac{1}{\sqrt{2}} ( \vert 1_{\zeta 1} 0_{\zeta 2} \rangle + (-1)^{n}  i\,\vert 0_{\zeta 1} 1_{\zeta 2} \rangle).           
\end{equation}

The vorticity field is better visualized in the optical field-strength space (sometimes referred as configuration-space-like space). In this representation $\mathcal{E}$ is the eigenvalue of the optical field-strength operator $\hat{\mathcal{E}}=(\hat{a} + \hat{a}^{\dag})/\sqrt{2}$, proportional to the first quadrature of the optical field, and analogously $\hat{\mathcal{P}}=-i (\hat{a} - \hat{a}^{\dag})/\sqrt{2}$ is related to the second quadrature \cite{Schleich2001}. Then, the normalized wavefunction corresponding to the quantum state heralded when a click in mode $3$ is measured is given by \cite{Agarwal2012}
\begin{equation}\label{Psi}
\Psi (\mathcal{E}_{1}, \mathcal{E}_{2})=\sqrt{ \frac{2}{\pi (1+\eta'^{2}) e^{4r} }} (\mathcal{E}_{1} + (-1)^{n} \,i \eta'\,\mathcal{E}_{2})\,e^{-\frac{\mathcal{E}_{1}^{2} + \mathcal{E}_{2}^{2}}{2 e^{2r}}},
\end{equation}
where we have chosen $\zeta$ real. Figures $\ref{F2}$ and $\ref{F3}$ show the probability ${P}=\vert \Psi  (\mathcal{E}_{1}, \mathcal{E}_{2}) \vert^{2}$ and phase ${\varphi}=\arg \{\Psi(\mathcal{E}_{1}, \mathcal{E}_{2})\}$ densities for a squeezing factor $r=0.3$ and different values of $\eta'$, namely a circular squeezed vortex for $\eta' = 1$ (Figure $\ref{F2}$) and an elliptical squeezed vortex for $\eta' = 5$ (Figure $\ref{F3}$). Comparing the probability densities it is easy to visualize the lack of symmetry as the value of $\eta'$ moves away from $1$. Likewise, comparing the phase densities we can see the appearance of a phase step as ellipticity increases.

It is important to outline here that the circular squeezed vortex is an eigenstate of the abstract angular momentum operator $\hat{L}_{z}=-i (\hat{\mathcal{E}}_{1}\partial_{\mathcal{E}_{2}} - \hat{\mathcal{E}}_{2}\partial_{\mathcal{E}_{1}})$ with eigenvalues $\pm 1$, so the vortex state carries an orbital angular momentum of $\pm\hbar$. This fact could be interesting for implementing specific CV-QIP protocols. In this respect, this operator can be implemented by means of a $3$ $dB$ DC with two photon number-resolving detectors (PNRDs) connected at its outputs \cite{Mattioli2015}. This is easily proved applying the operator which represents the DC, given by Equation (\ref{DC}) with $\theta(\omega_{s})=\pi/2$, on the abstract angular momentum operator above presented
\begin{equation}
\hat{L}_{z}' = \hat{B}_{1 2} (\pi/2)\, \hat{L}_{z} \, \hat{B}_{1 2}^{\dag} (\pi/2)=\hat{n}_{1}' - \hat{n}_{2}',
\end{equation}
where $\hat n _{j}'=\hat{a}_{j}'^{\dag} \hat{a}_{j}'$ is the photon number operator related to each mode $j$ and the prime denotes the output modes related to the DC. Therefore, measuring the difference of number of photons at each output $1'$ and $2'$, we can obtain the order and abstract handedness of the quantum vortex associated to the modes $1$ and $2$. This supports the use of this kind of states in a hybrid QIP, where DV information is transported by a CV channel, and recovered by this measurement technique. Moreover, it should be remarked that the use of SPADs in the measurement of the ancillary photons results in the conditional state to be mixed, since photodiodes do not resolve the number of photons, but under the conditions above outlined, specifically high transmittivity of the directional couplers and moderate squeezing, the resultant state is close to the ideal one \cite{Kim2008}.  This fact is fully prevented if SSPDs are used. However, if necessary, a model of inefficient detection which could be readily adapted to this scheme to fit real data from an experiment has been introduced in \cite{Ourjoumtsev2009}. 

\begin{figure}[t]
\centering
\includegraphics[width=0.47\textwidth]{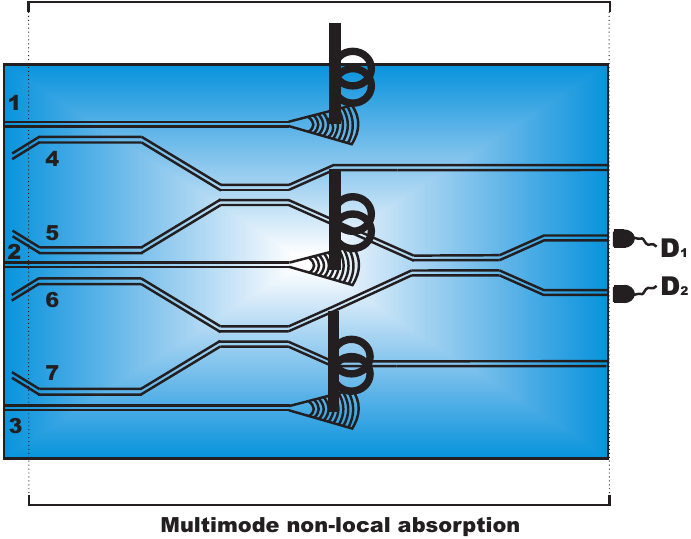}
\vspace {0cm}\,
\hspace{0cm}\caption{\label{F4}\small{(Color online) Sketch of the photonic circuit proposed for engineering three-mode CV entangled quantum states. A click in detectors $D_{1}$ or $D_{2}$ heralds the entangling of the three input states.}}
\end {figure}

Likewise, we would like to outline that higher Hilbert dimensions are accessible by this scheme via the nesting of periodically poled waveguides and $3$ $dB$ DCs. This is achieved by means of the multimode non-local absorption of one photon from a number of single-mode squeezed vacua excited in different waveguides. To illustrate, a three-mode CV entangled state could be produced by the generation of three squeezed vacua in the waveguides $1$, $2$ and $3$ in a similar way as that shown above (Figure \ref{F1}A),  and the absorption of photons by means of four weak DCs, as shown in Figure \ref{F4}. The quantum state after going through these couplers would be given by
\begin{align}\nonumber
\vert \Psi \rangle = (1-i\frac{r_{1}}{t_{1}} \hat{a}_{1} \hat{a}_{4}^{\dag})(1-i\frac{r_{2}}{\sqrt{2} t_{2}} \hat{a}_{2} (\hat{a}_{5}^{\dag}+\hat{a}_{6}^{\dag}))\\
(1-i\frac{r_{3}}{t_{3}} \hat{a}_{3} \hat{a}_{7}^{\dag}) \vert 0_{\zeta \,1} \, 0_{\zeta \, 2} \, 0_{\zeta \,3} \, 0_{4} \,0_{5} \,0_{6} \,0_{7} \rangle.
\end{align}
The photons coupled to the lossy channels find then a series of $3$ $dB$ DCs which erase the information about the source from which the photon is absorbed. The measurement of one photon in detector $D_{1}$ would herald the entangling of the three single-mode squeezed states in the following way
\begin{align}\nonumber
\langle 1_{D1}\vert \Psi \rangle &\propto \frac{r_{1}}{t_{1}}\vert 1_{\zeta \,1} \, 0_{\zeta \, 2} \, 0_{\zeta \,3} \rangle \\
&+ \frac{r_{2}}{t_{2}} e^{i\pi/4}\vert 0_{\zeta \,1} \, 1_{\zeta \, 2} \, 0_{\zeta \,3} \rangle+ \frac{r_{3}}{t_{3}} e^{i\pi/2}\vert 0_{\zeta \,1} \, 0_{\zeta \, 2} \, 1_{\zeta \,3} \rangle.
\end{align}
A similar outcome would be obtained detecting a photon with detector $D_{2}$. In this configuration, the entangled modes are collected into fibers by means of grating couplers. This scheme can be easily widened to a higher number of modes at the cost, however, of lowering the probability of generation of the state, since some of the absorbed photons are lost. Additionally, an adapted scheme to temporal multiplexing with heralding could be used to get access to ultra-large number of entangled modes \cite{Yokoyama2013}. 

It should be noted that the states thus produced are interesting in the context of continuous variable cluster states. Recently, it has been demonstrated that non-ideal Gaussian cluster states would present limitations in long computations \cite{Ohliger2010}, introducing the necessity of a resource of non-Gaussian states for measurement-based quantum computing \cite{Andersen2015}. This device could act as a generator of path-encoded non-Gaussian CV cluster states.

\subsection{DV quantum vortex states}

\begin{figure}[t]
\centering
\includegraphics[width=0.47\textwidth]{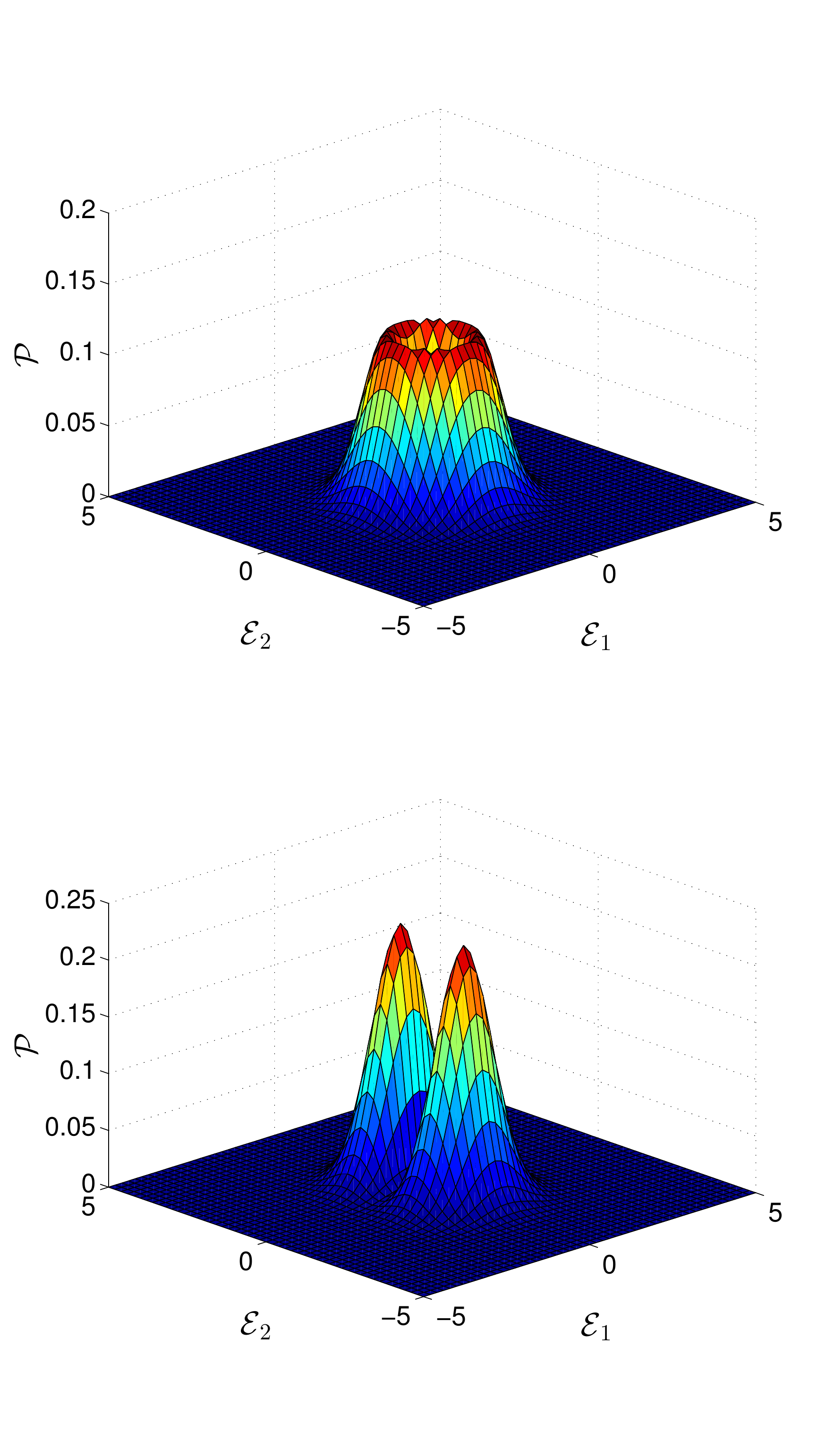}
\vspace {0cm}\,
\hspace{0cm}\caption{\label{F5}\small{(Color online) Plots of probability density of DV quantum vortex states with $\eta'=1$ (circular, upper figure) and $5$ (elliptical, lower figure).}}
\end {figure}

Additionally, it is interesting to note that this device could be also used for a second purpose. If instead of being interested on CV we move to DV-QIP, this design shows the ability to produce entanglement as well. This is directly carried out by lowering the power of the input pump, in such a way that two-photon states are generated at the waveguides $1$ and $2$. Taking the lower terms of the the single-mode squeezing operator given by Equation (\ref{SMSO}), we would have at each waveguide j the following quantum state \cite{Agarwal2012}
\begin{equation}
\vert 0_{\zeta\, j}\rangle \approx \frac{1}{\sqrt{\cosh(r)}} (\vert 0_{j} \rangle + \frac{e^{i\theta_{s}} \tanh(r)}{\sqrt{2}}   \, \vert 2_{j} \rangle),
\end{equation}
having a two-photon state with a certain probability. We have neglected here higher-order photon-number states since they will be produced with a very low probability in this regime. Substituting these states in Equation (\ref{GenState}), and following the same steps above outlined for the generation of CV quantum vortex states, the heralded output quantum states after detection of a single photon in modes $3$ or $4$ would be given by
\begin{subequations} \label{NSVS}
\begin{equation}\label{NSVSa}
\langle 1_{3} 0_{4} \vert \Psi_{D} \rangle \propto  \vert 1_{1} 0_{2} \rangle -\eta \,\tan(\phi_{2}/2) e^{-i\phi_{1}} \vert 0_{1} 1_{2} \rangle ,            
\end{equation}
\begin{equation}\label{NSVSb}
\langle 0_{3} 1_{4}\vert \Psi_{D} \rangle \propto \vert 1_{1} 0_{2} \rangle +\eta \,\cot(\phi_{2}/2) e^{-i\phi_{1}} \vert 0_{1} 1_{2} \rangle,
\end{equation}
\end{subequations}
in such a way that adjusting the phases $\phi_{1}$ and $\phi_{2}$ we have any state on the Bloch sphere, or a qubit. Choosing odd multiples of $\pi/2$ for $\phi_{1}$ we obtain again quantum states as these given by Equation (\ref{SVSE}) for $r=0$. These are non-squeezed elliptical vortex states. Likewise, setting the second phase shifter in the MZI in such a way that $\phi_{2}=2 \arctan(\eta^{-1})$ (or $\phi_{2}=2 \arctan(\eta) )$, we will have a circular non-squeezed quantum vortex, eigenstate of $\hat{L}_{z}$ as well. The abstract angular momentum is in this case $\pm \hbar$ per photon. Figure $\ref{F5}$ is devoted to show the probability densities corresponding to these non-squeezed vortex states for values of $\eta'=1$ and $5$ in order to compare with those shown in Figures $\ref{F2}$ and $\ref{F3}$. Note the higher localization of these states in the optical field-strength space reaching larger values of probability in comparison with their squeezed counterparts. The phase densities will be the same as those sketched in Figures $\ref{F2}$ and $\ref{F3}$. 

\subsection{Discussion}

Now, we address some of the merits that this scheme shows. Firstly, we would like to point out that a similar scheme has been proposed in bulk optics with polarization-encoded Schr\"{o}dinger kittens \cite{Ourjoumtsev2007b, Ourjoumtsev2009}. However, our integrated scheme presents noteworthy differences as well as new possibilities. Current technology has shown recently the ability to generate in few-centimeters-in-size chips bright squeezed light \cite{Yoshino2007} as well as both high-fidelity manipulation \cite{Carolan2015} and fast modulation \cite{Bonneau2012} of quantum states of light. We can estimate the performance of our device with experimentally available parameters. Propagation losses around $\approx 0.3$ $dB$ $cm^{-1}$ have been reported in a quantum relay chip in the telecommunication C band \cite{Martin2012}. Such advanced device showed geometrical losses of $\approx 1$ $dB$ due to the sinusoidal bends related to the DCs. Likewise, a SPDC photon flux as high as $1.4 \times 10^{7}$ pairs $nm^{-1}$ $mW^{-1}$ $s^{-1}$ in a $1$ $cm$-long PPLN waveguide operating in the telecom C and L bands has been recently reported \cite{Jin2014}. With these parameters, considering a typical value of $1$ $dB$ in the output coupling losses using micro-lenses systems (guide-to-fibre), SSPDs in the lossy channels with typical detection efficiencies of $\approx 10\%$ and a transmittivity of $99\%$ in the DCs related to the non-local absorption stage (Figure \ref{F1}C), a flux of $\approx 1.25 \times 10^{4}$ $nm^{-1}$ $mW^{-1}$ $s^{-1}$ entangled quantum states would be measured in a $5$ $cm$-long chip after the PPLN zone. It is important to outline that this figure could be improved by the optimization of the propagation efficiency (losses as low as $0.1$ $dB$ $cm^{-1}$ have been reported) and with the use of high-efficiency detectors, like the future superconducting nanowire-based single photon detectors \cite{Sahin2015}.

Likewise, guided modes of light have also proven to be robust carriers of quantum information over long distances and in various degrees of freedom \cite{Tanzilli2012}. Specifically, our approach is compact and polarization independent which, on the one hand, increases the stability and gets rid of insertion losses; and, on the other hand, allows long-distance transmission, since polarization maintaining fibers which show higher losses than standard fibers are not necessary in order to guide the output light. It also presents access to an universal set of states as well as error tolerance to the fabrication defects or asymmetries in the couplers responsible of the non-local absorption of photons, thanks to the reconfigurable circuit of stage 3, in addition to the capacity of dealing with higher-dimension quantum states in a reasonable size by means of nesting (scalability), issues difficult to manage with bulk optics circuits. Furthermore, our proposal shows the novel ability of generation of both CV and DV entangled quantum states in the same device by only changing the measurement configuration in a such a way that just a monolithic circuit can be applied on different QIP tasks. Another interesting feature of this scheme is an improvement of a factor of two in the probability of generation of the non-Gaussian states in comparison with other proposals where only one lossy channel is used. However, it could be argued that with this configuration the probability of success in the generation of DV entangled states is low in comparison with other current methods \cite{Jin2014}. This is the price to be paid in this hybrid configuration. This issue could be sorted out via the use of tunable electro-optic directional couplers \cite{Martin2012, Barral2015} in the second stage of the circuit (Figure \ref{F1}C), with the aim of acting as a $3$ $dB$ directional couplers and sending, with a probability of $50\%$ for each channel $1$ and $2$, one photon to the lossy channels $3$ or $4$, respectively. 

Additionally, we would like to outline that this scheme could be also used in entangled coherent state QIP. The squeezed vacuum and squeezed single photon states are also known as even and odd Schr\"{o}dinger kittens, respectively, due to the high fidelity they show with odd Schr\"{o}dinger cats for low values of squeezing \cite{Lund2005, Neergaard2006}. They can be written as
\begin{equation}
\vert 0_{\zeta} \rangle \propto \vert \alpha \rangle + \vert -\alpha \rangle, \qquad \vert 1_{\zeta} \rangle \propto \vert \alpha \rangle - \vert -\alpha \rangle.
\end{equation}
Using this representation in, for instance, Equation (\ref{SVSa}), we obtain
\begin{equation}\label{SVScatA}
\begin{split}
\langle 1_{3} 0_{4} \vert \Psi_{D} \rangle \propto\, &C_{-} ( \vert \alpha \rangle_{1}   \vert \alpha \rangle_{2} -  \vert -\alpha \rangle_{1} \vert -\alpha \rangle_{2}  ) \\
+  &C_{+} ( \vert \alpha \rangle_{1}  \vert -\alpha \rangle_{2} -  \vert -\alpha \rangle_{1} \vert \alpha \rangle_{2}),   
\end{split}   
\end{equation}
with $C_{\pm}=1 \pm \eta \tan(\phi_{2}/2) e^{-i\phi_{1}}$. It has been shown that two-mode entangled coherent states of this kind can perform quantum teleportation and rotations of coherent qubits \cite{Jeong2001, Ourjoumtsev2009}, opening the possibilities of this integrated device in coherent quantum communications and quantum computation through the tuning of the coefficients $C_{\pm}(\phi_{1}, \phi_{2})$. In fact, applying the scheme proposed in Figure $\ref{F4}$ in this representation, cluster-type entangled coherent states could be produced \cite{Munhoz2008}.

Finally, it is noted that the use of a directional coupler in the generation of a different family of quantum vortices has been dealt with in \cite{Bandyopadhyay2011}, but in that proposal neither the problem of generation of the non-Gaussian states necessary to produce vorticity nor the reconfigurability necessary for manipulation and the measurement of these states were covered. 

Now, in the next section we study the quantum features of the quantum states produced by this device.

\section{Non-classicality and entanglement}

We start this section studying the negativity of the Wigner function to assess qualitatively the non-classicality and of the quantum states generated with this scheme. We would like to outline that these features are related with the non-Gaussian nature of these states. As the relative phase does not affect the non-classicality and the entanglement of the quantum states presented above, we focus our attention on the quantum vortices without loss of generality. In order to simplify the calculation we start with the following ansatz 
\begin{equation}\label{Ansatz}
\vert L \rangle= \frac{1}{\sqrt{1+\eta'^{2}}} \{ \vert 1 0 \rangle + i \eta' \vert 0 1 \rangle \}.
\end{equation}
This state is an elliptical DV quantum vortex, equivalent to an elliptical CV quantum vortex given by Equation (\ref{SVSE}) if $\zeta=0$ is chosen. The Wigner function of this state can be readily worked out by the usual methods obtaining \cite{Furusawa2015}
\begin {equation}\label{W1}
\begin{split}
&W_{\vert L\rangle} (\alpha_{1}, \alpha_{2})= \frac{4}{\pi^{2}\,(1+\eta'^{2})} e^{-2(\vert \alpha_{1} \vert^{2} + \vert \alpha_{2} \vert^{2})}   \\
\{ (4\vert \alpha_{1} &\vert^{2} - 1) + \eta'^{2} ( 4\vert \alpha_{2} \vert^{2} - 1) - 8\eta' \vert \alpha_{1} \vert \vert \alpha_{2} \vert \sin(\delta_{1} - \delta_{2})  \}, 
\end{split}
\end{equation}
where $\alpha_{j}\equiv\vert \alpha_{j} \vert e^{i\delta{j}}=(\mathcal{E}_{j}+i\mathcal{P}_{j})$ with $j=1, 2$. This is clearly a non-Gaussian Wigner function. The regions of phase space with negative values of this function give a qualitative measure of the non-classicality of the state, since classical states are positive along the entire space. They are given by the following condition
\begin{equation}\label{N1}
\vert \alpha_{1} \vert^{2} + \eta'^{2} \vert \alpha_{2} \vert^{2} - 2\eta' \vert \alpha_{1} \vert \vert \alpha_{2} \vert \sin(\delta_{1} - \delta_{2}) < \frac{1+\eta'^{2}}{4}.
\end{equation} 
So this points out the non-classicality of the DV quantum vortices given by Equations (\ref{NSVS}) when $\phi_{1}=\pm\pi/2$, since there are a great number of values of $\alpha_{1}$, $\alpha_{2}$ and $\phi_{2}$ which fulfill the condition (\ref{N1}) for a given $\eta'$. For instance, in the case of a DV circular quantum vortex, the following condition is obtained
\begin{equation}\label{N2}
\vert {\alpha}_{1} \vert^{2} + \vert {\alpha}_{2} \vert^{2} - 2\vert \alpha_{1} \vert \vert \alpha_{2} \vert \sin(\delta_{1} - \delta_{2})< 1/2.
\end{equation}

Next, we pay attention on CV quantum vortices given by Equation (\ref{SVSE}). With the help of the following relation between density matrices related by a squeezed transformation \cite{Agarwal2012}
\begin{equation}
\hat{\tilde{\rho}}=\hat{S}(\zeta) \hat{\rho} \hat{S}^{\dag}(\zeta)  \rightarrow  W_{\hat{\tilde{\rho}}} (\alpha)=W_{\hat{\rho}}(\tilde{\alpha}),
\end{equation}
where $\tilde{\alpha}=\alpha \cosh(r) - \alpha^{*} e^{i\theta_{s}} \sinh(r)$, we can directly relate the Wigner function for the ansatz above calculated, Equation (\ref{W1}), with the Wigner function of the family of quantum states given by Equations (\ref{SVS}) by only exchanging $\alpha_{j}$ by $\tilde{\alpha}_{j}$ in that equation. This of course keeps the non-Gaussian shape of the Wigner function. Likewise, the negative regions are also obtained with the same substitution in Equation (\ref{N1}). For instance, setting $\zeta$ real we have
\begin{equation}\label{N3}
\vert \alpha_{1} \,e^{r} \vert^{2} + \eta'^{2} \vert \alpha_{2} \,e^{-r} \vert^{2} - 2\eta' \vert \alpha_{1} \vert \vert \alpha_{2} \vert \sin(\delta_{1} - \delta_{2}) < \frac{1+\eta'^{2}}{4}.
\end{equation} 

\begin{figure}[t]
\centering
\includegraphics[width=0.51\textwidth]{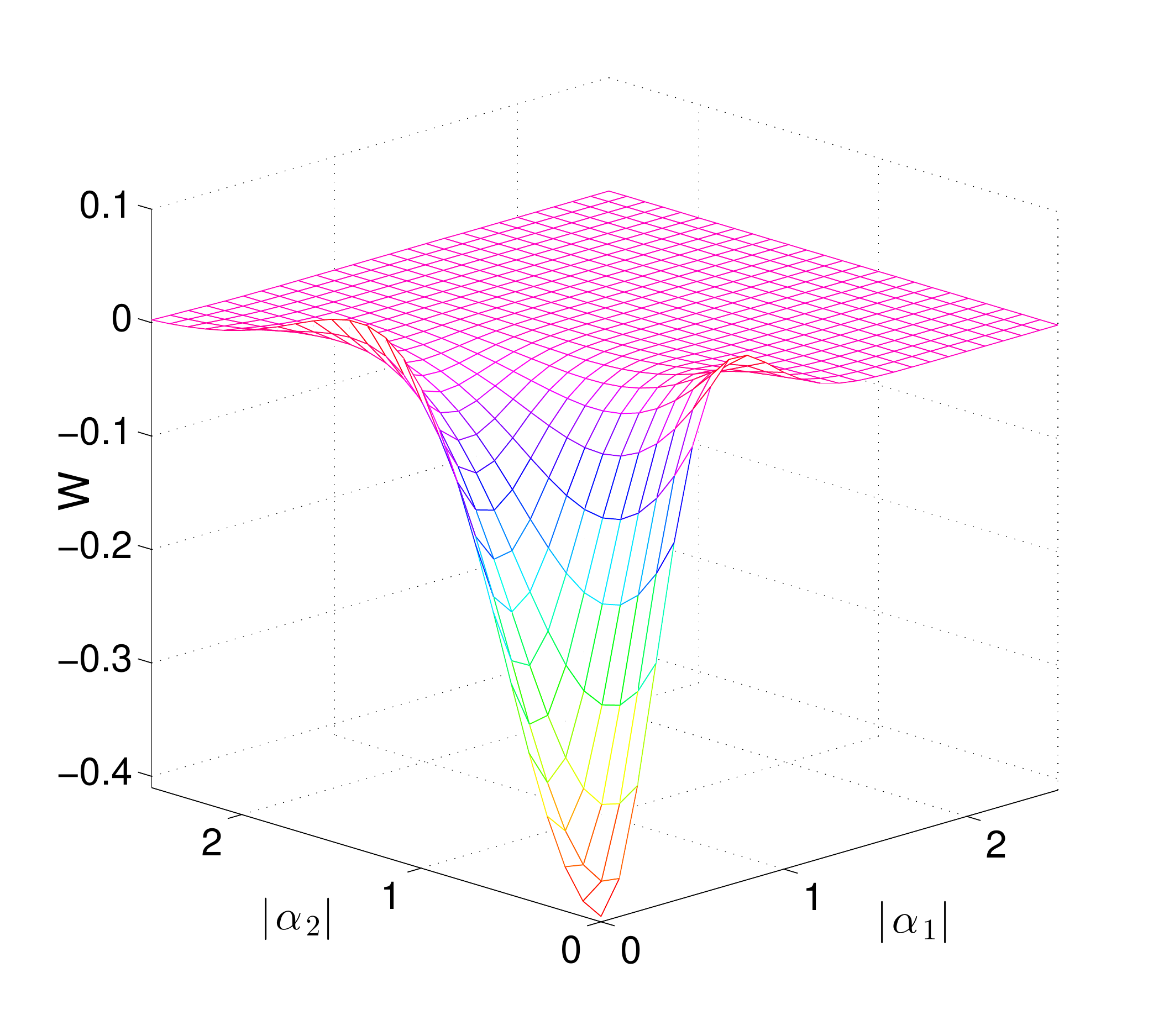}
\vspace {0cm}
\hspace{0cm}\caption{\label{F6}\small{(Color online) Plot of the Wigner function related to a CV circular quantum vortex with $r=0.3$. We have set $\delta_{1}=\pi/2$ and $\delta_{2}=0$ in order to represent it in two dimensions.}}
\end {figure}

It is shown in Figure \ref{F6} a plot of the Wigner function corresponding to a circular CV vortex state ($\eta'=1$) with a real squeezing factor with value $r=0.3$ and where we have set $\delta_{1}=\pi/2$ and $\delta_{2}=0$ in order to be able to sketch it in two dimensions \cite{Seshadreesan2015}. It is easy to see that it takes on negative values along a large area of the phase space, as given by Equation (\ref{N3}).



On the other hand, we are interested on a quantitative measure of entanglement. To this end we choose here the logarithmic negativity because of its appealing properties. Namely, that it is computable, additive and provides an upper bound on the efficiency of distillation, so it measures to which extent a quantum state is useful for a certain QIP protocol or how many resources are needed to produce it. This is defined by \cite{Vidal2002}
\begin{equation}\label{LogNeg}
E=\log_{2}(1+2\mathcal{N}),
\end{equation}
with $\mathcal{N}$ the modulus of the sum of all negative eigenvalues of the partial transpose of the density matrix associated to the state to be measured. Furthermore, for pure states Equation (\ref{LogNeg}) can be written as
\begin{equation}\label{LogNeg2}
E=2\,\log_{2}\left(\sum_{\alpha} c_{\alpha}\right),
\end{equation}
where $c_{\alpha}$ are the Schmidt coefficients of the pure state in a suitable orthonormal basis $e_{\alpha}^{(j)}$, with $j=1, 2$ \cite{Vidal2002}. For a given squeezing, the maximum CV entangled vortex states will be the circular ones given by Equation (\ref{SVSC}). With the help of the circular basis
\begin{equation}
\hat{b}_{1}=\frac{\hat{a}_{1}-i\,e^{i n\pi} \,\hat{a}_{2}}{\sqrt{2}}, \qquad \hat{b}_{2}=\frac{\hat{a}_{1}+i \,e^{i n\pi}\, \hat{a}_{2}}{\sqrt{2}},
\end{equation}
we can easily work out the Schmidt decomposition of the circular quantum vortex given in Equation (\ref{SVSC}). In this basis the state can be written as
\begin{equation}
\vert \Psi \rangle = \hat{S}_{1 2}(\zeta)\, \hat{b}_{1}^{\dag} \,\vert 0_{1} \,0_{2} \rangle,
\end{equation}
where $\hat{S}_{1 2}(\zeta)= \exp\,\{\zeta \hat{b}_{1}^{\dag} \hat{b}_{2}^{\dag}-\zeta^{*} \hat{b}_{1} \hat{b}_{2}\}$ is the two-mode squeezing operator. Applying the decomposition of the two-mode squeezing operator for $\zeta=r$ (Appendix A), we can write the state in the circular Fock basis as
\begin{equation}
\vert \Psi\rangle= \sum_{n=0}^{\infty} c_{n}(r) \vert n+1, n\rangle,
\end{equation}
with $c_{n}(r)=\sqrt{n+1} \,\tanh^{n}(r)/\cosh^{2}(r)$ the corresponding Schmidt coefficients. Setting these values on Equation (\ref{LogNeg2}), we have
\begin{equation}\label{LogNeg3}
E=2\,\log_{2} \left[ \cosh^{-2}(r)\,(1+ \sum_{n=1}^{\infty}\sqrt{n+1} \tanh^{n}(r)\,)\right].
\end{equation}
So entanglement quickly increases with squeezing and, in the limit of no squeezing ($r=0$), the state would be fully untangled ($E=0$), since it would be equivalent to not having quantum light within the chip at all. It is interesting to note that the same logarithmic negativity has been obtained studying the generalized vortex state produced by means of photon subtraction from a two-mode squeezed vacuum \cite{Agarwal2011}. It should be outlined however, that these states present different quantum features. For instance, they are not eigenstates of $\hat{L}_{z}$, as could be readily shown by calculating the expected value of the abstract angular momentum.

Generalizing the above result to CV elliptical vortex states, $E$ will decrease as the ellipticity $\eta'$ tends to zero in the following way
\small{\begin{equation}\label{LogNeg4}
\begin{split}
&E=2\,\log_{2} \left(\sum_{n} \vert c_{n} (r, \Phi) \vert \right) = \\
&2\,\log_{2} \left[\cosh^{-2}(r\sin(2\Phi))(1+ \sum_{n=1}^{\infty} \sqrt{n+1} \vert \tanh^{n}(r\sin(2\Phi)) \vert )   \right],
\end{split}
\end{equation}}
where we have defined $\Phi=\arctan(\eta')$ and $\vert c_{n} (r, \Phi) \vert$ are the Schmidt coefficients, which recover the circular vortex state for $\Phi=(2l+1)\pi/4$ ($\vert c_{n} (r, (2l+1)\pi/4)\vert \equiv c_{n} (r)$), with $l$ any integer. A detailed calculation of this Equation is shown in Appendix A. On the other hand, unlike CV quantum vortices, DV vortex states keeps the value of logarithmic negativity equal to $1$ for any $\Phi$ different of $\pi/2$ \cite{Agarwal2012}, since there are always an eigenbasis which diagonalizes the state (Equation (\ref{EllipBasis}) with $\Phi=\phi_{2}/2$).


\begin{figure}[t]
\centering
\includegraphics[width=0.49\textwidth]{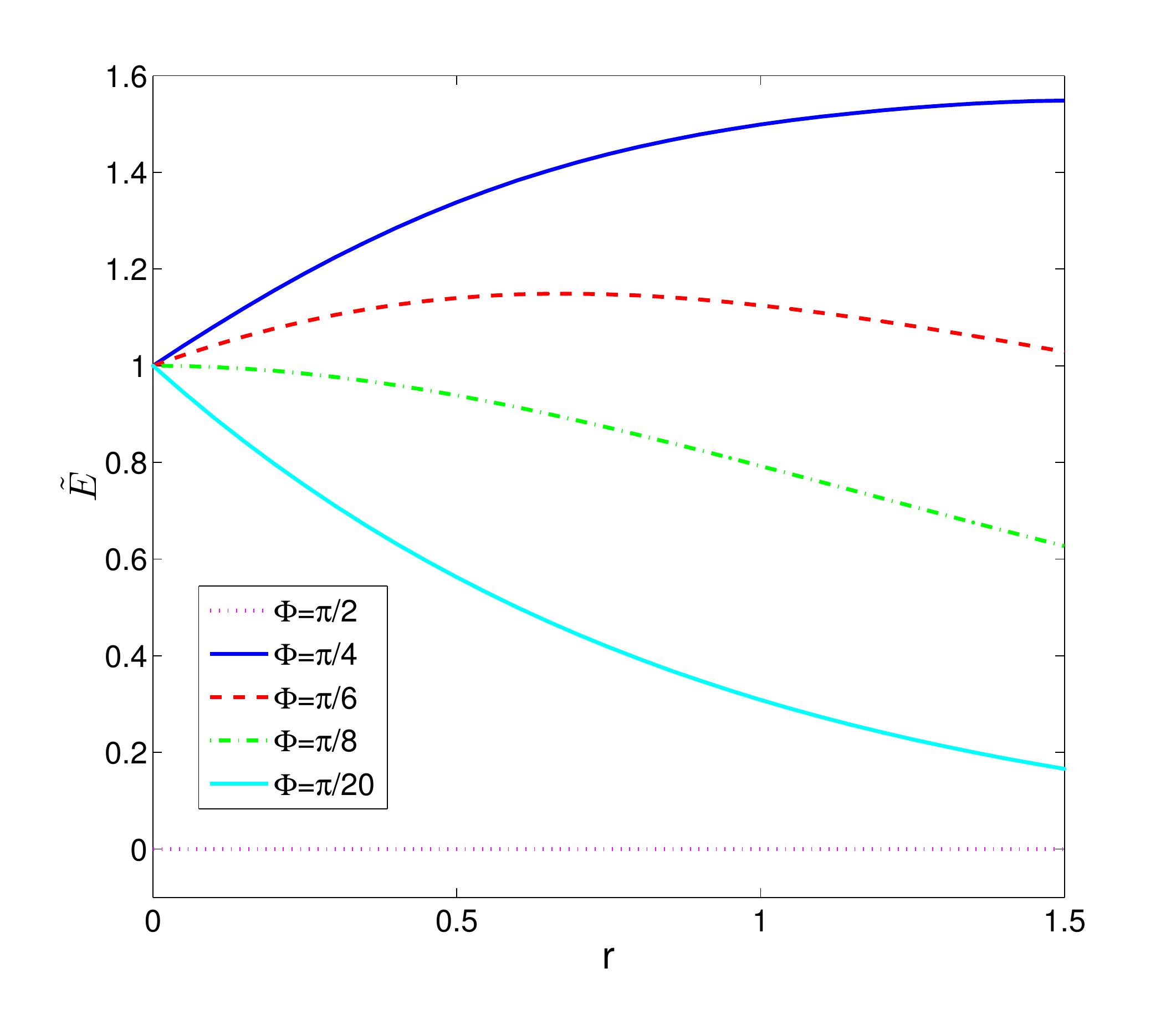}
\vspace {0cm}\,
\hspace{0cm}\caption{\label{F7}\small{(Color online) Ratio between the logarithmic negativity of CV vortex states and that corresponding to a two-mode squeezed vacuum over a range of values of the squeezing parameter $r$ and different values of ellipticity given by $\Phi$.}}
\end {figure}

Therefore, the most interesting states in terms of entanglement presented here are the CV quantum vortices. In order to visualize the degree of entanglement these states reach, we use the ratio between the logarithmic negativity of the quantum vortices and that corresponding to a two-mode squeezed vacuum, $2\,\log_{2} (e^{r})$, given by the following relation \cite{Agarwal2011} 
\begin{equation}
\tilde{{E}}=\left(\sum_{n} c_{n}(r, \Phi)\,e^{-r} \right)^{2}.
\end{equation}
In Figure $\ref{F7}$ this ratio is shown over a range of values of $r$ and for various values of ellipticity $\Phi$ (or equally $\eta'$). It is noted that $\tilde{E}>1$ for $\Phi=\pi/4$ (solid, blue), in such a way that the circular vortex state entanglement increases faster and higher with squeezing than that related to the two-mode squeezed vacuum. Even for values different but close to $\pi/4$ the state keeps a strong entanglement for moderate values of squeezing (dashed, red). As we get closer to values of $\Phi$ multiples of $\pi$, the state becomes linearized, or experiences a lost of symmetry (Figure \ref{F2} {vs} Figure \ref{F3}), and the entanglement diminishes with $r$, such that $\tilde{E} \rightarrow 0$ (dot-dashed, green \& solid, cyan). In the limit of $\Phi=\pi/2$ the state is fully untangled and $\tilde{E}=0$ (dot, magenta). It is interesting to note that we observe an inverse relationship between generalized quantum polarization in the optical field-strength space and entanglement \cite{Linares2011, Barral2013}. The lack of symmetry in this space increases the polarization degree and decreases the entanglement.



\section{Conclusions}

In this article we have proposed a monolithic quantum photonic circuit for QLE via a reconfigurable superposition of photon subtraction on two single-mode squeezed states produced with periodically poled waveguides, directional couplers and phase shifters. This scheme allows the generation of both CV and DV strongly entangled quantum states of light in general, and quantum vortices in particular, on the same chip, modifying only the power of the input pump from which the quantum states are generated. We have  demonstrated the ability of nesting that this design presents, increasing the number of dimensions of CV entanglement and thus showing a great potential in future integrated applications for QIP. Moreover, we have studied the application of this device to the production, manipulation and detection of photonic quantum vortex states, showing a measurement scheme of vorticity order and abstract handedness with possibilities in QIP. We have also studied relevant quantum properties like non-Gaussianity, non-classicality and entanglement, via the optical field-strength probability distributions, Wigner function and logarithmic negativity. Specifically, we have shown this scheme produces large values of entanglement for moderate values of squeezing.

\section*{Acknowledgements}
Authors wish to acknowledge the partial financial support of this work by 
Ministerio de Econom{\'i}a y Competitividad, Central Government of Spain, Contract number FIS2013-46584-C2-1-R, and Fondo Europeo de Desenvolvemento Rexional 2007-2013 (FEDER). 

\section*{Appendix A}

With the help of the following transformation
\begin{equation} \label{EllipBasis}
 \begin{pmatrix}
 \hat{b}_{1} \\
 \hat{b}_{2}
 \end{pmatrix}
 =
  \begin{pmatrix}
     \cos(\Phi) &  \,-i\,e^{i n \pi}\sin(\Phi) \\
     \sin(\Phi) &  i\,e^{i n \pi} \cos(\Phi)
  \end{pmatrix} 
    \begin{pmatrix}
 \hat{a}_{1} \\
 \hat{a}_{2}
  \end{pmatrix} , 
\end{equation}
where $\Phi=\arctan(\eta')$, we can readily rewrite Equation (\ref{SVSE}) in the new basis as
\begin{equation}\label{SVSE2}
\vert \Psi \rangle = \hat{S}_{1}(r \cos(2\Phi)) \, \hat{S}_{2}(- r \cos(2\Phi)) \, \hat{S}_{1 2} (r \sin(2\Phi))\,\vert 1_{1} 0_{2} \rangle.
\end{equation}
The single-mode squeezing operators $\hat{S}_{1}$ and $\hat{S}_{2}$ in the previous Equation do not contribute to entanglement since they can be cancelled out by local unitary operations on each mode \cite{Kim2002}, so they can be neglected in this calculation onwards. Applying the decomposition theorem of the two-mode squeezing operator
\begin{equation}
\hat{S}_{1 2}(r)=e^{\tanh(r)\, \hat{b}_{1}^{\dag} \hat{b}_{2}^{\dag}} \,e^{-\ln \cosh(r) (\hat{b}_{1}^{\dag}\hat{b}_{1} + \hat{b}_{2}^{\dag}\hat{b}_{2}+1)} \,e^{-\tanh(r)\, \hat{b}_{1} \hat{b}_{2}},
\end{equation}
we obtain the following state
\begin{equation}\label{SVSE3}
\vert \Psi' \rangle =  \sum_{n=0}^{\infty} c_{n} (r, \Phi) \vert n+1, n\rangle,
\end{equation}
with $c_{n} (r, \Phi)=\sqrt{n+1} \tanh^{n}(r\sin(2\Phi))/\cosh^{2}(r\sin(2\Phi))$ and where we have applied the local operations above discussed, that is $\vert \Psi' \rangle =  \hat{S}_{1}(-r \cos(2\Phi)) \, \hat{S}_{2}(r \cos(2\Phi))\, \vert \Psi \rangle$. Since these coefficients can show negative values, we calculate the negativity $\mathcal{N}$ of this state, given by $\mathcal{N}=\vert \sum_{k} \lambda_{k} \vert$, with $\lambda_{k}$ the negative $k$ eigenvalues of the partial transpose of the density operator $\hat{\rho}^{PT}$. From Equation (\ref{SVSE3}) the partial transpose is
\begin{equation}\label{PT}
\hat{\rho}^{PT}=\sum_{n, m=0}^{\infty} c_{n}(r, \Phi) c_{m}(r, \Phi)\,\vert n+1, m \rangle \langle m+1, n \vert. 
\end{equation}
We can diagonalize the non-positive terms ($n \neq m$) of this matrix by means of the following eigenvectors
\begin{equation}
\begin{split}
&(\vert n+1, m\rangle +  \vert m+1, n \rangle)( \langle n+1, m\vert +  \langle m+1, n \vert), \\
&(\vert n+1, m\rangle -  \vert m+1, n \rangle)( \langle n+1, m\vert -  \langle m+1, n \vert),
\end{split}
\end{equation}
with eigenvalues $(\frac{c_{n}c_{m}}{2}, -\frac{c_{n}c_{m}}{2})$, respectively. Therefore the negativity is given by 
\begin{equation}
\mathcal{N}=\frac{1}{2}\vert \sum_{n \neq m} c_{n}(r, \Phi) \,c_{m}(r, \Phi)\vert,
\end{equation}
and the logarithmic negativity (Equation (\ref{LogNeg})) is
\begin{equation}
E=\log_{2} \left( 1+\vert \sum_{n\neq m} c_{n}(r, \Phi) \,c_{m}(r, \Phi)\vert \right).
\end{equation}
By taking into account that $\sum_{n} c_{n}^{2}(r, \Phi)=1$, we finally obtain
\begin{equation}
E=2\,\log_{2}\left(\sum_{n} \vert c_{n}(r, \Phi) \vert \right).
\end{equation}


\begin{thebibliography}{70}
\bibitem{Braunstein2005} S.L. Braunstein and P. van Loock {\it Rev. Mod. Physics} {\bf 77}, 513 (2005).
\bibitem{Eisert2002} J. Eisert, S. Scheel, and M. B. Plenio {\it Phys. Rev. Lett.} {\bf 89},
137903 (2002).
\bibitem{Fiurasek2002} J. Fiurasek {\it Phys. Rev. Lett.} {\bf 89}, 137904 (2002).
\bibitem{Giedke2002} G. Giedke and J.I. Cirac {\it Phys. Rev. A} {\bf 66}, 032316 (2002).
\bibitem{Takahashi2010} H. Takahashi, J.S. Neergaard-Nielsen, M. Takeuchi, M. Takeoka, K. Hayasaka, A. Furusawa and M. Sasaki {\it Nature Photonics} {\bf 4}, 178 (2010).
\bibitem{Kim2008} M.S. Kim {\it J. Phys. B: At. Mol. Opt. Phys.} {\bf{41}}, 133001 (2008).
\bibitem{Agarwal1991} G.S. Agarwal and K. Tara {\it Phys. Rev. A} {\bf 43}, 492 (1991).
\bibitem{Zavatta2004} A. Zavatta, S. Viciani, and M. Bellini, {\it Science} {\bf 306}, 660 (2004).
\bibitem{Dakna1997} M. Dakna, T. Anhut, T. Opartny, L. Kn\"{o}ll and D.-G. Welsch  {\it Phys. Rev. A} {\bf 55} (4), 3184 (1997).
\bibitem{Neergaard2006} J. S. Neergaard-Nielsen, B. Melholt Nielsen, C. Hettich, K. M{/o}lmer and E. S. Polzik {\it Phys. Rev. Lett.} {97}, 083604 (2006).
\bibitem{Ourjoumtsev2007} A. Ourjoumtsev, H. Jeong, R. Tualle-Brouri and P. Grangier {\it Nature} {\bf 448}, 784 (2007). 
\bibitem{Kim2008b} M.S. Kim, H. Jeong, A. Zavatta, V. Parigi and M. Bellini {\it Phys. Rev. Lett.} {\bf 101}, 260401 (2008).
\bibitem{Zavatta2009} A. Zavatta, V. Parigi, M.S. Kim, H. Jeong and M. Bellini {\it Phys. Rev. Lett.} {\bf 103}, 140406 (2009).
\bibitem{GarciaPatron2004} R. Garc{\'i}a-Patr{\'o}n, J. Fiurasek, N.J. Cerf, J. Wenger, R. Tualle-Brouri and P. Grangier {\it Phys. Rev. Lett.} {\bf 93}, 130409 (2004).
\bibitem{Nha2004} H. Nha and H.J. Carmichael {\it Phys. Rev. Lett.} {\bf 93}, 020401
(2004).
\bibitem{Lee2011} N. Lee, H. Benichi, Y. Takeno, S. Takeda, J. Webb, E. Huntington and A. Furusawa, {\it Science} {\bf 332}, 330 (2011).

\bibitem{Molina2007} G. Molina-Terriza, J.P. Torres and L. Torner {\it Nature Phys.} {\bf 3}, 305 (2007).
\bibitem{Agarwal1997} G.S. Agarwal, R.R. Puri, and R.P. Singh {\it Phys. Rev. A} {\bf 56}, 4207 (1997).
\bibitem{Agarwal2006} G.S. Agarwal and J. Banerji {\it J. Phys. A} {\bf  39}, 11503 (2006).
\bibitem{Bandyopadhyay2011} A. Bandyopadhyay and R. P. Singh {\it Opt. Commun.} {\bf 284}, 256 (2011).
\bibitem{Bandyopadhyay2011b} A. Bandyopadhyay, S. Prabhakar and R. P. Singh {\it Phys. Lett. A} {\bf 375}, 1926 (2011).
\bibitem{Agarwal2011} G.S. Agarwal {\it New J. Phys.} {\bf 13}, 073008 (2011).
\bibitem{Banerji2014} A. Banerji, R.P. Singh and A Bandyopadhayay {\it Opt. Commun.} {\bf 330}, 85 (2014).
\bibitem{Zhu2012} K. Zhu, S. Li, X. Zheng and H. Tang {\it J. Opt. Soc. Am. B} {\bf 29} (6), 1179 (2012).
\bibitem{Zhu2012b} K. Zhu, S. Li, Y. Tang, X. Zheng and H. Tang {\it Chin. Phys. B} {\bf 21} (8), 084204 (2012).
\bibitem{Li2015} Y. Li, F. Jia, H. Zhang, J. Huang and L. Hu {\it Laser Phys. Lett.} {\bf 12}, 115203 (2015). 
\bibitem{Luis2013} A. Luis and A.S. Sanz {\it Phys. Rev. A} {\bf 87}, 063844 (2013).
\bibitem{Luis2015} A. Luis and A.S. Sanz {\it Phys. Rev. A} {\bf 92}, 023832 (2015).

\bibitem{Tanzilli2012} S. Tanzilli, A. Martin, F. Kaiser, M.P. De Micheli, O. Alibart and D.B. Ostrowsky {\it Laser \& Photon. Rev.} {\bf 6} (1), 115 (2012). 
\bibitem{Andersen2015} U.L. Andersen, J.S. Neergaard-Nielsen, P. van Loock and A. Furusawa {\it Nature Phys.} {\bf 11}, 713 (2015).
\bibitem{OBrien2009} J.L. O'Brien, A. Furusawa and J. Vukovic {\it Nature Photon.} {\bf 3}, 687 (2009).
\bibitem{Silberhorn2007} C. Silberhorn {\it Contemporary Physics} {\bf 48} (3), 143 (2007).
\bibitem{Rogers2015} S. Rogers, X. Lu, W.C. Jiang and Q. Lin {\it Appl. Phys. Lett.} {\bf 1070} (4), 041102 (2015).
\bibitem{Dutt2015} A. Dutt, K. Luke, S. Manipatruni, A.L. Gaeta, P. Nussenzveig and M. Lipson {Phys. Rev. Applied} {\bf 3}, 044005 (2015).
\bibitem{Jin2014} H. Jin, F.M. Liu, P. Xu, J.L. Xia, M.L. Zhong, Y. Yuan, J.W. Zhou, Y.X. Gong, W. Wang, and S.N. Zhu {\it Phys. Rev. Lett.} {\bf 113}, 103601 (2014).
\bibitem{Masada2015} G. Masada, K. Miyata, A. Politi, T. Hashimoto, J.L. O'Brien and A. Furusawa {\it Nature Photon.} {\bf 9}, 316 (2015).
\bibitem{Carolan2015} J. Carolan, C. Harrold, C. Sparrow, E. Mart{\'\i}n-L{\'o}pez, N.J. Russell, J.W. Silverstone, P.J. Shadbolt, N. Matsuda, M. Oguma, M. Itoh, G. Marshall, M.G. Thompson, J.C.F Matthews, T. Hashimoto, J.L. O'Brien and A. Laing {\it Science} {\bf 349} (6249), 711 (2015).
\bibitem{Sahin2015} D. Sahin, A. Gaggero, J.-W. Weber, I. Agafonov, M.A. Verheijen, F. Mattioli, J. Beetz, M. Kamp, S. H\"{o}fling, M.C.M. van de Sanden, R. Leoni, and A. Fiore {\it IEEE J. Sel. Top. Quantum Electron.} {\bf 21}, 1 (2015).
\bibitem{Najafi2015} F. Najafi, J. Mower, N.C. Harris, F. Bellei, A. Dane, C. Lee, X. Hu, P. Kharel, F. Marsili, S. Assefa, K.K. Berggren and D. Englund {\it Nat. Commun.}  {\bf 6}, 5873 (2015).


\bibitem{Tanzilli2000} S. Tanzilli, H. de Riedmatten, W. Tittel, H. Zbinden, P. Baldi, M. de Micheli, D.B. Ostrowsky and N. Gisin {\it Electron. Lett.} {\bf 37}, 26 (2000).
\bibitem{Suhara2009} T. Suhara {\it Laser \& Photon. Rev.} {\bf 3} (4), 370 (2009).
\bibitem{Yoshino2007} K. Yoshino, T. Aoki and A. Furusawa {\it Appl. Phys. Lett.} {\bf 90}, 041111 (2007).
\bibitem{Saleh2009} M. Saleh, B.A.E. Saleh and M.C. Teich {\it Phys. Rev. A} {\bf 79}, 053842 (2009).
\bibitem{Lugani2010} J. Lugani, S. Ghosh and K. Thyagarajan {\it Phys. Rev. A} {\bf 83}, 062333 (2011).
\bibitem{Linares2008} J. Li\~nares, M.C. Nistal and D. Barral {\it New J. Phys.} {\bf 10}, 063023 (2008).
\bibitem{Barral2015a} D. Barral and J. Li\~nares {\it J. Opt. Soc. Am. B} {\bf 32} (9), 1993 (2015).
\bibitem{Bai2012} N. Bai, E. Ip, Y.-K. Huang, E. Mateo, F. Yaman, M.-J. Li, S. Bickham, S. Ten, J. Li\~nares, C. Montero, V. Moreno, X. Prieto, V. Tse, K. M. Chung, A. P. T. Lau, H.-Y. Tam, C. Lu, Y. Luo, G.-D. Peng, G. Li, and T. Wang {\it Opt. Express} {\bf 20}, 2668 (2012).
\bibitem{Kanter2002} G.S. Kanter, P. Kumar, R.V. Roussev, J. Kurz, K.R. Parameswaran and M.M. Fejer {\it Opt. Express} {\bf 10} (3), 177 (2002).
\bibitem{Campos1989} R.A. Campos, B.A.E. Saleh and M.C. Teich {\it Phys. Rev. A} {\bf 40} (3), 1371 (1989).
\bibitem{Martin2012} A. Martin, O. Alibart, M.P. De Micheli, D.B. Ostrowsky and S. Tanzilli {\it New J. Phys.} {\bf 14} 025002 (2012).
\bibitem{Bonneau2012} D. Bonneau, M. Lobino, P. Jiang, C.M. Natarajan, M.G. Tanner, R.H. Hadfield, S.N. Dorenbos, V. Zwiller, M.G. Thompson, and J.L. O'Brien {\it Phys. Rev. Lett. }{\bf 108}, 053601 (2012).
\bibitem{Shadbolt2012} P.J. Shadbolt, M.R. Verde, A. Peruzzo, A. Politi, A. Laing, M. Lobino, J.C.F. Matthews, M.G. Thompson and J.L. O'Brien {\it Nat. Photonics} {\bf 6}, 45 (2012)
\bibitem{Barral2015} D. Barral, M.G. Thompson and J. Li\~nares {\it J. Opt. Soc. Am. B} {\bf 32} (6), 1165 (2015).
\bibitem{Eisaman2011} M.D. Eisaman, J.F.A Migdall and S.V. Polyakov {\it Rev. Sci. Instrum.} {\bf 82}, 071101 (2011).
\bibitem{Agarwal2012} G.S. Agarwal {\it Quantum optics} (Cambridge University Press, Cambridge, 2012).
\bibitem{Schleich2001} W.P. Schleich {\it Quantum optics in phase space} (Wiley-VCH, Weinheim, 2001).
\bibitem{Mattioli2015} F. Mattioli, Z. Zhou, A. Gaggero, R. Gaudio, S. Jahanmirinejad, D. Sahin, F. Marsili, R. Leoni and A. Fiore {\it Supercond. Sci. Technol.} {\bf 28}, 104001 (2015).
\bibitem{Ourjoumtsev2009} A. Ourjoumtsev, F. Ferreyrol, R. Tualle-Brouri and P. Grangier {\it Nature Phys.} {\bf 5}, 189 (2009).
\bibitem{Yokoyama2013} S. Yokoyama, R. Ukai, S.C. Armstrong, C Sornphiphatphong, T. Kaji, S. Suzuki, J. Yoshinawa, H. Yonezawa, N.C. Meniucci and A. Furusawa {\it Nat. Phot.} {\bf 7}, 982 (2013).
\bibitem{Ohliger2010} M. Ohliger, K. Kieling and J. Eisert {\it Phys. Rev. A} {\bf 82}, 042336 (2010).
\bibitem{Ourjoumtsev2007b} A. Ourjoumtsev, A. Dantan, R. Tualle-Brouri and P. Grangier {\it Phys. Rev. Lett.} {\bf 98}, 030502 (2007).
\bibitem{Lund2005} A.P. Lund and T.C. Ralph {\it Phys. Rev. A} {\bf 71}, 032305 (2005).
\bibitem{Jeong2001} H. Jeong, M.S. Kim and J. Lee {\it Phys. Rev. A} {\bf 64}, 052308 (2001).
\bibitem{Munhoz2008} P.P. Munhoz, F.L. Semiao, A. Vidiella-Barranco and J.A. Roversi {\it Phys. Lett. A} {\bf 372}, 3580 (2008).
\bibitem{Furusawa2015} A. Furusawa {\it Quantum states of light} (Springer, Tokyo, 2015).
\bibitem{Seshadreesan2015} K.P. Seshadreesan, J.P. Dowling and G.S. Agarwal {\it Phys. Scr.} {\bf 90}, 074029 (2015).
\bibitem{Vidal2002} G. Vidal and R.F. Werner {\it Phys. Rev. A} {\bf 65}, 032314 (2002).
\bibitem{Linares2011} J. Li\~nares, D. Barral, M.C. Nistal and V. Moreno {\it J. Mod. Opt.} {\bf 58}, 711 (2011).
\bibitem{Barral2013} D. Barral, J. Li\~nares and M.C. Nistal {\it J. Mod. Opt.} {\bf 60} (12), 941 (2013).
\bibitem{Kim2002} M.S. Kim, W. Son, V. Buzek and P.L. Knight {\it Phys. Rev. A} {\bf 65}, 032323 (2002).

\end{thebibliography}
\end{document}